\documentclass{appolb}
\usepackage{epsfig}

\usepackage{cite}
\newcommand{\as}{\alpha_{\rm s}}
\def\MSbar{\overline{\mathrm{MS}}}
\def\ep{\epsilon}
\def\calO{{\cal O}}

\begin{document}
\title{The QCD form factor of massive quarks \\
and applications
\thanks{
  Presented at {\it Matter To The Deepest: Recent Developments In Physics of
    Fundamental Interactions}, September 11-16 , 2009, Ustron, Poland.
}%
}
\author{S. Moch
\address{
Deutsches Elektronensynchrotron DESY \\
Platanenallee 6, D--15738 Zeuthen, Germany
}
}
\maketitle
\begin{abstract}
We review the electromagnetic form factor of heavy quarks with emphasis on 
the QCD radiative corrections at two-loop order in the perturbative expansion.
We discuss important properties of the heavy-quark form factor 
such as its exponentiation in the high-energy limit and its role 
in QCD factorization theorems for massive $n$-parton amplitudes.
\end{abstract}

%
%
\section{Introduction}
\label{sec:introduction}
The electromagnetic form factor of heavy quarks comprises the simplest example
of a scattering amplitude in Quantum Chromodynamics (QCD) 
for the study of mass effects in QCD hard scattering processes 
including radiative corrections at higher orders.
As such, it has received considerable attention in recent years. 

Explicit results for the two-loop QCD corrections have been obtained 
for the vector- and axial-vector coupling 
as well as the anomaly contributions~\cite{Bernreuther:2004ih,Bernreuther:2004th,Bernreuther:2005rw}.
Moreover, universal features of higher order radiative corrections, 
such as the exponentiation of logarithms in the heavy-quark mass $m$ together with
the infrared (soft) singularities within dimensional regularization 
have been addressed~\cite{Mitov:2006xs}.
As a consequence, these factorization properties in the soft and collinear limit imply 
for on-shell massive $n$-parton QCD amplitudes 
a simple relation between massless and massive scattering amplitudes 
at any order in perturbation theory~\cite{Mitov:2006xs,Becher:2007cu}.

In the following, we review briefly recent progress in studies of QCD corrections 
to the heavy-quark form factor.

%
%
\section{QCD radiative corrections}
\label{sec:higherorders}
The object of our interest are the scalar functions ${\cal F}_1$ and ${\cal F}_2$
which parametrize the vector current of an on-shell heavy-quark pair of momenta $k_1$
and $k_2$ and of mass $m$. 
We denote the associated vertex function as $\Gamma_\mu$ and 
for a photon of (space-like) virtuality $Q^2=-q^2 > 0$, we can write 
\begin{eqnarray}
\label{eq:ffdef}
{\lefteqn{
\Gamma_\mu(k_1,k_2)  \: = \: }}
\\
& &
{\rm i} e_{\rm q}\,
{\bar \psi}(k_1)\,
\left(
\gamma_{\mu\,}\, {\cal F}_1 (Q^2,m^2,\alpha_{s})
-
{{\rm i} \over 2 m}
\,
\sigma_{\mu \nu\,} q^\nu \, {\cal F}_2 (Q^2,m^2,\alpha_{s}) \right) \psi(k_2)
\, .
\nonumber
\end{eqnarray}
Here $e_{\rm q}$ is the charge and 
${\cal F}_1$ and ${\cal F}_2$ are the electric and magnetic space-like quark form factors.
They are gauge invariant, but in perturbative QCD at higher orders in general divergent 
and their power expansion in the strong coupling $\alpha_{s}$ at the scale $\mu^2$ reads,
\begin{eqnarray}
\label{eq:aexp}
  {\cal F}(\alpha_{s}) 
&=&  \sum\limits_{i=0}^{\infty} 
 \left(
{\alpha_{s}(\mu^2) \over 4\pi}
\right)^i
\, {\cal F}^{(i)}
\, \equiv \, \sum\limits_{i=0}^{\infty} \left(a_s(\mu^2)\right)^{\,i}\, {\cal F}^{(i)}\, ,
\end{eqnarray}
where we have introduced the shorthand notation $a_s \equiv \alpha_{s} / (4\pi)$.

Up to the two-loop level there is only a relatively small number of Feynman diagrams to be computed. 
All divergences are regulated in $D=4-2\epsilon$ dimensions and the results
are given as a Laurent expansion in $\epsilon$.
Except for additional diagrams from heavy-quark self-energy corrections on
the external lines, which contribute to the wave function renormalization, 
the diagrams are in one-to-one correspondence to the calculation 
of the form factor of massless quarks (see e.g. Ref.~\cite{Moch:2005id}).

The reduction of all Feynman integrals to so-called master integrals proceeds in a standard way. 
Since the computation of the QCD corrections in Eq.~(\ref{eq:aexp}) is a
one-scale problem (dependent on the ratio $Q^2/m^2$) 
the master integrals can be expressed in terms of harmonic polylogarithms (HPLs)~\cite{Remiddi:1999ew}.
For the two-loop corrections to $\calO (\epsilon)$ HPLs up to weight $w=5$ are needed, 
$H_{m_1,...,m_w}(x)$, $m_j = 0,\pm 1$, which depend on the conformal variable $x$,  
\begin{equation}
  \label{eq:spacelike-xdef}
  \displaystyle
  x = {\sqrt{Q^2 + 4m^2} - \sqrt{Q^2} \over \sqrt{Q^2 + 4m^2} + \sqrt{Q^2}} 
  \, ,
\end{equation}
with $Q^2 > 0$ and $0 \le x \le 1$. 
The analytic computation of the master integrals has used the method of
differential equations together with two independent numerical checks at selected phase space
points based either on sector decomposition or the evaluation of a Mellin-Barnes representation.

The ultraviolet divergences arising in the computation of the Feynman diagrams do require renormalization 
of the strong coupling constant $\alpha_{s}$, the external (heavy-quark) wave function $\psi$ 
and the heavy-quark mass $m$ (typically taken to be the pole mass).
The renormalized form factors ${\cal F}_1$ and ${\cal F}_2$ 
are obtained by adding the appropriate counter-terms $CT_1$ and $CT_2$.
The renormalization of the bare quantities is performed multiplicatively with
all necessary constants 
in the $\MSbar$-scheme (respectively in the on-shell scheme for $m$) being known~\cite{Bernreuther:2004ih,Melnikov:2000qh,Melnikov:2000zc}.

The two-loop corrections to ${\cal F}_1$ and ${\cal F}_2$ 
to $\calO (\epsilon^0)$ have been obtained in~\cite{Bernreuther:2004ih} 
and all terms of $\calO (\epsilon)$ are new results of~\cite{Gluza:2009yy}.
Up to order $\calO (\epsilon^0)$ at two-loops full agreement between~\cite{Bernreuther:2004ih} 
and the first independent check~\cite{Gluza:2009yy} has been established.

For space-like kinematics the results for ${\cal F}_1$ and ${\cal F}_2$ are real 
whereas for time-like kinematics above production threshold ($Q^2 > 4 m^2$) 
both ${\cal F}_1$ and ${\cal F}_2$ develop an imaginary part, 
\begin{equation}
  \label{eq:tmlkdef}
  {\cal F}_i \,=\,
  \Re {\cal F}_i 
  + {\rm i}\, \Im {\cal F}_i 
  \, ,
\end{equation}
which can be obtained by means of a suitable (complex) continuation of $Q^2$, 
giving rise to $x \to -x$ in Eq.~(\ref{eq:spacelike-xdef}) and all arguments of HPLs.

%
%
\section{Exponentiation}
\label{sec:exponentiation}
The exponentiation of the heavy-quark form factor is based on the universality of soft and collinear radiation 
and the respective singular terms in the electric form factor ${\cal F}_1$, 
i.e. the (soft gluon) poles in $\ep$ and the logarithms in the heavy-quark mass. 
The logarithms emerge in the (space-like) high-energy limit, $Q^2 \gg m^2$, i.e. $x \to 0$ 
in Eq.~(\ref{eq:spacelike-xdef}), as 
\begin{eqnarray}
  \label{eq:L-def}
  L = \ln\left( {Q^2\over m^2}\right)\, .
\end{eqnarray}
The magnetic form factor ${\cal F}_2$ is power suppressed by $m^2/Q^2$ at high energies.
The large logarithms of Sudakov type~\cite{Sudakov:1954sw} in Eq.(\ref{eq:L-def}) can be resummed.
More generally ${\cal F}_1$ fulfills the following evolution equation, see e.g.~\cite{Collins:1980ih,Mitov:2006xs},
\begin{eqnarray}
  \label{eq:ffdeq}
- \mu^2 {\partial \over \partial \mu^2} \ln {\cal F}_1\left({Q^2 \over \mu^2},{m^2 \over \mu^2},\alpha_{s},\ep\right) 
\!&=&\!
  {1 \over 2} K\left({m^2 \over \mu^2},\alpha_{s},\ep \right)
+ {1 \over 2} G\left({Q^2 \over \mu^2},\alpha_{s},\ep \right) \, ,
\,\,\,\,\,\,
\end{eqnarray}
where the dependence on the various scales has been made explicit. 
QCD factorization at the scale $\mu$ allows to separate to logarithmic accuracy in the limit $Q^2 \gg m^2$ 
the dependence on the hard scale $Q$ (associated with the function $G$)
from that on the heavy-quark mass $m$ (resting in the function $K$).
Both functions, $G$ and $K$, are subject to renormalization
group equations~\cite{Collins:1980ih} governed by the same 
(well-known) cusp anomalous dimension $A$~\cite{Kodaira:1982nh,Moch:2004pa,Vogt:2004mw},
\begin{eqnarray}
  \label{eq:KGdeq}
  - \lim_{m \to 0} \mu^2 {d \over d \mu^2} K\left({m^2 \over \mu^2},\alpha_{s},\ep \right) 
  \, = \,
  \mu^2 {d \over d \mu^2} G\left({Q^2 \over \mu^2},\alpha_{s},\ep \right) 
  \, = \,  A(\alpha_{s}) \, .
\end{eqnarray}
In dimensional regularization, the solution of the evolution
equation~(\ref{eq:ffdeq}) proceeds analogous to the construction for the
massless form factor~\cite{Moch:2005id,Magnea:1990zb,Moch:2005tm,Dixon:2008gr}.
However, in contrast to the massless case where $K$ is a pure infrared counter term 
(dependent on the cusp anomalous dimension $A$), 
the heavy-quark mass in Eq.~(\ref{eq:ffdeq}) acts as an additional regulator in the collinear limit.
Hence the infrared sector differs and $K$ takes particular values in the
perturbative expansion, see~\cite{Mitov:2006xs,Gluza:2009yy}.

The (ultraviolet) renormalized form factor $\ln {\cal F}_1$ 
reads in space-like kinematics as a function of the renormalized coupling $\alpha_{s}(\mu^2)$ 
\begin{eqnarray}
\label{eq:massFFexp}
{\lefteqn{
    \ln \, {\cal F}_1 \left({Q^2 \over \mu^2},{m^2 \over \mu^2},a_{s}(\mu^2),\ep \right)
    \, = \,
     }}
\\ &&
  {1 \over 2}
  \int\limits_0^{Q^2/\mu^2} {d \xi \over \xi}
  \Bigg\{
  G\left(\bar a(\xi \mu^2)\right) 
  +
  K\left(\bar a(\xi \mu^2 m^2/Q^2)\right) 
  -
  \int\limits_{\xi m^2/Q^2}^\xi {d \lambda \over \lambda}
  A\left({\bar a}(\lambda \mu^2) \right)
  \Bigg\}
\, .
\nonumber
\end{eqnarray}
On the right hand side all quantities are defined in $D$ dimensions 
as functions of the $D$-dimensional strong coupling ${\bar a}$ 
(to be expanded in terms of the ordinary coupling $\alpha_{s}$ in four dimensions).
The exponentiation needs an additional integration constant $C({\bar a}(\mu^2,\ep),\ep)$, 
which is determined through matching to fixed-order results in perturbation theory.
The perturbative coefficients of the function $G$ agree completely with those of the form factor for massless quarks.
In the latter case the available three-loop information~\cite{Moch:2005id,Moch:2005tm,Baikov:2009bg} 
allows to make partial predictions of the high-energy expansion of ${\cal F}_1$ 
based on Eq.~(\ref{eq:massFFexp}) beyond the two-loop result of~\cite{Bernreuther:2004ih,Gluza:2009yy}.

%
%
\section{QCD amplitudes}
\label{sec:amplitudes}
For on-shell amplitudes of $n$-parton processes the factorization ansatz 
has to be extended in comparison to Eq.~(\ref{eq:ffdeq}).
In $D$-dimensions a generic $n$-parton amplitude (massless and massive alike) 
can be expressed as a product of functions ${\cal J}_{\rm p}$, ${\cal S}_{\rm p}$ and ${\cal H}_{\rm p}$ 
each governing different regions of kinematics (see~\cite{Catani:1998bh,Sterman:2002qn,MertAybat:2006mz}).
\begin{eqnarray}
\label{eq:QCDfacamplitude}
{\lefteqn{
| {\cal M}_{\rm p}\rangle \, = }}
\\
& &
{\cal J}_{\rm p}\left({Q^2 \over \mu^2},\as(\mu^2),\ep \right)
{\cal S}_{\rm p}\left(\{ k_i \},{Q^2 \over \mu^2},\as(\mu^2),\ep \right)
| {\cal H}_{\rm p} \rangle
\, ,
\nonumber
\end{eqnarray}
where the ket-notation $| {\cal M}_{\rm p}\rangle$ implies vectors (and matrices) in color space.
The jet function ${\cal J}_{\rm p}$ collects all collinear contributions of
the proccess $\rm p$. 
It is color-diagonal and given as the product of functions ${\cal J}_{[i]}$ 
for each external parton $i$. The functions ${\cal J}_{[i]}$ are conventionally chosen
to be the square root of the respective (massless or massive) form factor ${\cal F}_{1}$.
Soft radiation arising from the overall color flow is summarized by
the soft function ${\cal S}_{\rm p}$, which is a matrix in color space 
and the short-distance dynamics of the hard scattering are contained 
in the hard function ${\cal H}_{\rm p}$, which is to leading order proportional to the Born amplitude.

Eq.~(\ref{eq:QCDfacamplitude}) leads to a 
remarkably simple relation between $n$-parton amplitudes 
with massless and massive external partons in the small-mass limit.
If $m$ is small compared to all scales the entire non-trivial
mass dependence resides in the respective expression for the ${\cal J}_{[i]}$ (i.e. the form factor)
so that QCD factorization gives~\cite{Mitov:2006xs,Becher:2007cu} 
(see for a review~\cite{Moch:2007pj}),
\begin{eqnarray}
\label{eq:Mm-M0}
{\cal M}_{\rm p}^{(m)} &=&
\prod_{i\in\ \{{\rm all}\ {\rm legs}\}}\,
  \left(
    Z^{(m\vert0)}_{[i]}
  \right)^{1 \over 2}\,
  \times\
{\cal M}_{\rm p}^{(m=0)}\, ,
\end{eqnarray}
where we the upper index denotes (non-)zero parton masses. 
The universal and manifestly process-independent multiplicative function $Z^{(m\vert 0)}_{[i]}$ 
is directly given in terms of ratios of the respective form factors,
\begin{eqnarray}
\label{eq:Z}
{\lefteqn{
Z^{(m\vert0)}_{[i]}\left({m^2 \over \mu^2},\as,\ep \right)
\, = \,}}
\\
& &
{\cal F}_{[i]}^{(m)}\left({Q^2\over \mu^2},{m^2\over\mu^2},\as,\ep \right)
\left({\cal F}_{[i]}^{(m=0)}\left({Q^2\over \mu^2},\as,\ep \right)\right)^{-1}
\, ,
\nonumber
\end{eqnarray}
where the index $i$ denotes the (massive) parton and $\as = \as(\mu^2)$.
$Z^{(m\vert0)}_{[i]}$ is a function of the (process-independent) ratio of
scales $\mu^2/m^2$ as the (process-dependent) scale $Q$ cancels completely in the ratio.
This definition however, 
excludes terms with explicit dependence on the number of heavy quarks 
in Eq.~(\ref{eq:Z}) (and, therefore in Eq.~(\ref{eq:Mm-M0}) as well), 
because at two-loops (and beyond) additional process-dependent terms enter in their description~\cite{Becher:2007cu}.

Within QCD Eq.~(\ref{eq:Mm-M0}) has been applied in the computation of the (virtual) 
two-loop QCD corrections to hadronic top-quark pair-production~\cite{Czakon:2007ej,Czakon:2007wk} 
in the limit when all Mandelstam invariants are large, $s, |t|, |u| \gg m^2$. 
Recent advances beyond the small mass limit have been concerned with the exact
evaluation of Eq.~(\ref{eq:QCDfacamplitude}) for all kinematics. 
To that end, in particular the soft function ${\cal S}_{\rm p}$ and the
structure of its associated anomalous dimensions $\Gamma_{\rm p}$ have been
subject of intense investigation. 
For the latter quantity $\Gamma_{\rm p}$ all-order expressions in the case of massless~\cite{Becher:2009cu,Gardi:2009qi} 
and massive partons~\cite{Becher:2009kw} have been suggested and 
explicit results for the two-loop soft anomalous dimension of massive partons
have appeared 
recently~\cite{Beneke:2009rj,Czakon:2009zw,Ferroglia:2009ep}.
Based on Eq.~(\ref{eq:Mm-M0}) this knowledge has been applied to derive the singularity structure of the two-loop virtual amplitude 
for top-quark hadro-production with complete kinematics dependence~\cite{Ferroglia:2009ii} 
which was found to be in agreement with the small mass limit~\cite{Czakon:2007ej,Czakon:2007wk}.

%
%
\section{Conclusions}
\label{sec:conclusions}

We have provided a brief review of the known perturbative QCD corrections to the heavy-quark form factor. 
Thanks to~\cite{Bernreuther:2004ih} both the electric and the magnetic form factor 
${\cal F}_1$ and ${\cal F}_2$ are known to two-loop order, an important result which has been recently confirmed 
and extended to higher order in $\epsilon$ by~\cite{Gluza:2009yy}.
The heavy-quark form factor will also be part of future calculations of QCD
radiative corrections to $n$-parton scattering processes at two-loop order.

It is a long known fact that the logarithms in the heavy-quark mass $m$ in the 
form factor do exponentiate.
These logarithms emerge in the limit of small masses upon neglecting all terms $m^2 \ll Q^2$ 
and within dimensional regularization the exponentiation extends to the complete 
singularity structure, i.e. the soft poles in $\ep$.
The exponentiation is governed by the cusp anomalous dimension $A$ as well as
the functions $K$ and $G$. 
Strikingly, the latter is identical for the case of massless and massive quarks, while the
infrared sector ($K$) of course differs.

The knowledge of the complete singularity structure of scattering amplitudes and, 
in particular the factorization of amplitudes in soft and collinear limits has numerous applications. 
The factorization property may even be used as a tool in practical calculations,
a prominent example being the computation of the virtual corrections 
to heavy-quark hadro-production at two loops in QCD.

%
%

\end{document}